\documentclass[trans]{IEEEtran}
\usepackage{mathtools}
\usepackage{graphicx}
\usepackage[linesnumbered,ruled]{algorithm2e}
\SetKwFor{ParFor}{parallel for}{do}{end}
\SetKwRepeat{Do}{do}{while}
\usepackage{amsmath}
\usepackage{multicol,caption}
\usepackage{color}
\usepackage{amssymb}
\usepackage{bbm}
\newenvironment{Figure}
{\par\medskip\noindent\minipage{\linewidth}}
{\endminipage\par\medskip}

\begin{document}
	
	\onecolumn

	\title{Three-dimensional Cooperative Localization of Commercial-Off-The-Shelf Sensors}
	
	\author{
		\IEEEauthorblockN{
			Yulong Wang\IEEEauthorrefmark{1}\IEEEauthorrefmark{2},
			Shenghong Li\IEEEauthorrefmark{1},
			Wei Ni\IEEEauthorrefmark{1},
			David Abbott\IEEEauthorrefmark{1},
			Mark Johnson\IEEEauthorrefmark{1},
			Guangyu Pei\IEEEauthorrefmark{3},
			and Mark Hedley\IEEEauthorrefmark{1},
		}
		\IEEEauthorblockA{\IEEEauthorrefmark{1}CSIRO, Marsfield, Australia}\\
		\IEEEauthorblockA{\IEEEauthorrefmark{2}School of Computer Science (National Pilot Software Engineering School), Beijing University of Posts and Telecommunications, Beijing, China}\\
		\IEEEauthorblockA{\IEEEauthorrefmark{3}Boeing, Seattle, USA}
	}
	
	\markboth{Submitted to TASE}%
	{Shell \MakeLowercase{\textit{et al.}}: Bare Demo of IEEEtran.cls for IEEE Transactions on Magnetics Journals}

	\IEEEtitleabstractindextext{
		\begin{abstract}
			Many location-based services use Received Signal Strength (RSS) measurements due to their universal availability. In this paper, we study the association of a large number of low-cost Internet-of-Things (IoT) sensors and their possible installation locations, which can enable various sensing and automation-related applications. We propose an efficient approach to solve the corresponding permutation combinatorial optimization problem, which integrates continuous space cooperative localization and permutation space likelihood ascent search. A convex relaxation-based optimization is designed to estimate the coarse locations of blindfolded devices in continuous 3D spaces, which are then projected to the feasible permutation space. An efficient Cram\'er-Rao Lower Bound based likelihood ascent search algorithm is proposed to refine the solution. Extensive experiments were conducted to evaluate the performance of the proposed approach, which show that the proposed approach significantly outperforms state-of-the-art combinatorial optimization algorithms and achieves close-to-100\% accuracy with affordable execution time.
		\end{abstract}
		
}
	
	\maketitle
	
	\IEEEdisplaynontitleabstractindextext
	
	\IEEEpeerreviewmaketitle
	
	\begin{multicols}{2}
		\setlength{\parskip}{0.002pt}
		\section{Introduction}
		
		The Internet-of-Things (IoT) has been increasingly deployed in manufacturing, transportation, agriculture, and logistics for data collection and response automation~\cite{6740844,7879243}.
		While a vast amount of informative IoT data is being generated daily (if not hourly)~\cite{Xinchen2020WirelMag,Ni2021ComMag, Ni2013JSAC}, the provenance of the data is equally critical, if not more. An indispensable aspect of data provenance is the location at which the data is captured. In many situations, the IoT data is only meaningful if correctly associated with the three-dimensional (3D) locations of the origin of the data, especially in many public safety and security related IoT applications~\cite{Diya2021IOTJ, Bo2020TIFS}. One of these application scenarios is inside an airplane, where a large number of IoT sensors can be deployed to tag and track safety devices and equipment. Any missing or misplaced devices and equipment could lead to catastrophic consequences, for example, missing life vests. With a correct location associated with every piece of sensory data, situational awareness can be delivered for the safety and control automation of an aircraft cabin.
		
		\subsection{Scenario, Motivation, and State of the Art}
		In this paper, we design a novel and practical cooperative localization technique which is able to associate a large number of low-cost, commercial-off-the-shelf (COTS) IoT sensors (and their captured data) correctly with the possible locations (i.e., installation points) of the sensors and report misplaced sensors, e.g., inside an airplane cabin. Our cooperative localization technique is based on the received signal strengths (RSSs) between the sensors, due to the universal availability of RSS measurements, the cost-effectiveness of COTS IoT devices, and the stringent regulative certification of any new custom-designed devices and systems in aircraft.
		
		The scenario of interest is very challenging in the sense that the RSS measurements typically provide very coarse ranging accuracy due to multiplicative errors. In contrast, other ranging measurements, such as Time-of-Arrival (TOA)~\cite{Savvides:2001:DFL:381677.381693, 878533, 1458289,Fan2021TCom}, Time-Difference-of-Arrival (TDOA)~\cite{1583910,Shenghong2019SysJ}, and Angle-of-Arrival (AOA)~\cite{4068140,Zhipeng2021TWC,Zhipeng2020TCom,Zhipeng2021JSAC}, could provide far better accuracy (provided the bandwidth is sufficiently wide). However, those ranging methods would require specialized hardware with substantially higher costs and, as a consequence, would not scale well~\cite{4533654,Yulong2021TVT,Yulong2020VTC}.
		
		The scenario of interest is computationally demanding, due to its combinatorial programming nature. Consider a practical passenger airplane with $N_t$ COTS IoT devices and $N_t$ installation locations. Each of the IoT devices is fitted arbitrarily at one of the $N_t$ locations. The association of the $N_t$ IoT devices and the $N_t$ 3D locations is in essence to build up a one-to-one mapping which is known to be NP-complete and mathematically intractable, e.g., Traveling Salesperson Problem (TSP)~\cite{Miller:1960:IPF:321043.321046} and Generalized Assignment Problem (GAP)~\cite{CATTRYSSE1992260, doi:10.3138/infor.45.3.123,Shmoys1993}.
		
		RSS measurements have been used for fingerprinting~\cite{4343996} and trilateration~\cite{1458287}, where the position of a blindfolded device is estimated by matching against the RSS signatures at surveyed locations or based on empirical path loss models. A survey is laborious, preventing the fingerprinting techniques from generalization into unsurveyed areas~\cite{4343996}. Empirical path
		loss models exhibit intense multiplicative noise, with standard deviation of up to 7 or 8 dB, and dramatically compromise the accuracy of trilateration~\cite{1458287}.
		
		Cooperative localization was proposed where all RSS measurements between blindfolded wireless devices, and between blindfolded devices and reference devices, are utilized to jointly estimate the locations of all the blindfolded devices~\cite{5762798}. However, a direct application of cooperative localization to the scenario of interest cannot achieve meaningful results in the presence of the aforementioned discrete known installation locations, due to the mathematical intractability of (mixed) integer programming. Moreover, RSS-based cooperative localization does not provide good resolution in 3D spaces with a high density of blindfolded devices, and can easily get confused by different possible elevations of a device. An error caused by the confusion can propagate throughout the network, due to the cooperative nature of the technique~\cite{1458287}. To the best of our knowledge, there has been to date no effective (and reliable) RSS-based 3D cooperative localization of a densely deployed array of COTS IoT devices in 3D spaces like an airplane cabin.

		\subsection{Contribution and Organization}
		
		This paper presents a new, efficient, RSS-based 3D cooperative localization technique in challenging and practically important scenarios, and achieves an accuracy which has not been achieved otherwise in the literature. Our technique locates and associates a large number of COTS IoT devices (and their sensory data) with known installation locations, by designing a joint optimization using continuous space cooperative localization and permutation space likelihood ascent search.
		
		The key contributions of the paper are as follows.
		\begin{itemize}
			\item A convex relaxation-based optimization is designed to jointly estimate the coarse locations of blindfolded devices in continuous 3D spaces. The results are projected to the discrete feasible solution space, i.e., a permutation of known 3D installation positions by extending the Kuhn-Munkres bipartite graph matching algorithm.
			\item An efficient likelihood ascent search algorithm is proposed to refine the blindfolded device assigned to every installed point, within a 3D region around the installed point. The 3D region is specified by the Cram\'er-Rao Lower Bound, hence reducing the searching complexity and relieving the prohibitive complexity of combinatorial optimization.
			\item An experimental platform based on Cypress CYW54907 system-on-chips is prototyped, and an efficient control protocol is designed to collect the pairwise RSS measurements between a large number of devices for localization.			
		\end{itemize}
		Extensive experiments including both large-scale simulations and lab tests with COTS WiFi devices are conducted to evaluate the proposed approach in terms of accuracy and execution time. The results show that the proposed approach significantly outperforms state-of-the-art combinatorial optimization algorithms and achieves close-to-100\% accuracy with affordable computations.
		
		The rest of this paper is organized as follows. Section \ref{sec:our_approach} firstly describes the system model and problem formulation. Section \ref{sec:algorithm} describes the proposed approach, including the continuous space cooperative localization algorithm, the modified Kuhn-Munkres algorithm for permutation-based feasible solution projection, and the general local search algorithm. In section~\ref{sec:CRB_LAS}, we propose the CRLB-based likelihood ascent search algorithm to accelerate the process of local search. Section \ref{sec:simulation} provides experimental results. Conclusions are drawn in Section \ref{sec:conclusion}.
		
		\section{Problem Formulation}
		\label{sec:our_approach}
		Let $N_{t}$ denote the number of blindfolded wireless devices to be located, each of which occupies one of $N_t$ possible installation locations that are known \textit{a prior} and given by $\{\mathring{\mathbf{p}}_i\}_{i=1}^{N_t}$, where $\mathring{\mathbf{p}}_i\triangleq[\mathring{x}_i, \mathring{y}_i, \mathring{z}_i]$.
		Additionally, let $N_a$ denote the number of wireless anchor devices deployed at known locations.
		Without loss of generality, we assume that device $i$ is to be located if $1\le i\le N_t$, or is an anchor if $N_t < i\le N_t+N_a$. The position of each wireless device is denoted by
		$\mathbf{p}_i=[x_i, y_i, z_i]$, ($i=1,...,N_t+N_a$).
		
		Each wireless device measures the RSS from all other devices. The RSS measured by device $i$ for wireless signals from device $j$ can be modeled as~\cite{LI2018253}
		\begin{equation}
		\label{eqn:rss_measure}
		\tilde{R}_{ij} = R_{ij} + \epsilon,
		\end{equation}
		where
		\begin{equation}
		\label{eqn:rss_ideal}
		R_{ij} = P_0-10\gamma \log_{10}\frac{\|\mathbf{p}_i-\mathbf{p}_j\|}{d_0} ~\text{(dBm)}
		\end{equation}
		denotes the ideal RSS in free space, $P_0$ is the received signal strength at the reference distance $d_0$ (meters), $\gamma$ is the path loss exponent, $\|\mathbf{p}_i-\mathbf{p}_j\|$ is the Euclidean distance between devices $i$ and $j$, $\epsilon \sim \mathcal{N}(0, \sigma^2)$ is the RSS measurement error caused by blockage, scattering, and reflection of radio signal.
		
		The locations of the blindfolded devices are estimated by solving the following constrained Maximum Likelihood Estimation (MLE) problem:
		\begin{align}
		\label{eqn:prob_model}
		\max_{\pi}& \prod_{i=1}^{N_t} \prod_{j=i+1}^{N_t+N_a} \mathcal{N}\Big(\tilde{R}_{ij}; R_{ij} , \sigma^2\Big)\\
		\label{eqn:pos_constraint}
		\text{s.t.} ~~ & \{\mathbf{p}_{\pi[i]}\}_{i=1}^{N_t} = \{\mathring{\mathbf{p}}_i\}_{i=1}^{N_t},
		\end{align}
		which is equivalent to
		\begin{align}
		\nonumber
		\mathcal{P}_1: ~~~ \min_{\pi} \quad c\big(\pi\big)\\
		\nonumber
		\text{s.t.} ~~ & (\ref{eqn:pos_constraint}),
		\end{align}
		with the cost function $c\big(\pi\big)$ defined by
		\begin{align}
		\label{eqn:cost_function}
		c\big(\pi\big)\triangleq \sum_{i=1}^{N_t} \sum_{j=i+1}^{N_t+N_a} \Big( P_0-10\gamma \log_{10} \frac{\|\mathbf{p}_{\pi[i]}-\mathbf{p}_{\pi[j]}\|}{d_0} - \tilde{R}_{ij} \Big)^2.
		\end{align}
		This problem falls into the category of permutation-based Combinatorial Optimization Programs (COP), and it is NP-hard due to above cost function~\cite{Garey:1990:CIG:574848}. Conventional algorithms for solving permutation COPs are metaheuristics, including genetic algorithm, tabu searching and simulated annealing. However, the conventional metaheuristics without considering the characteristic $\mathcal{P}_1$ will not perform well, as will be shown in section \ref{sec:simulation}. In the following section, we propose an efficient algorithm to solve $\mathcal{P}_1$, which achieves high accuracy in locating blindfolded wireless devices with affordable execution time.
		
		\section{The Proposed Approach}
		\label{sec:algorithm}
		
\begin{figure*}
			\centering
			\includegraphics[width=5in]{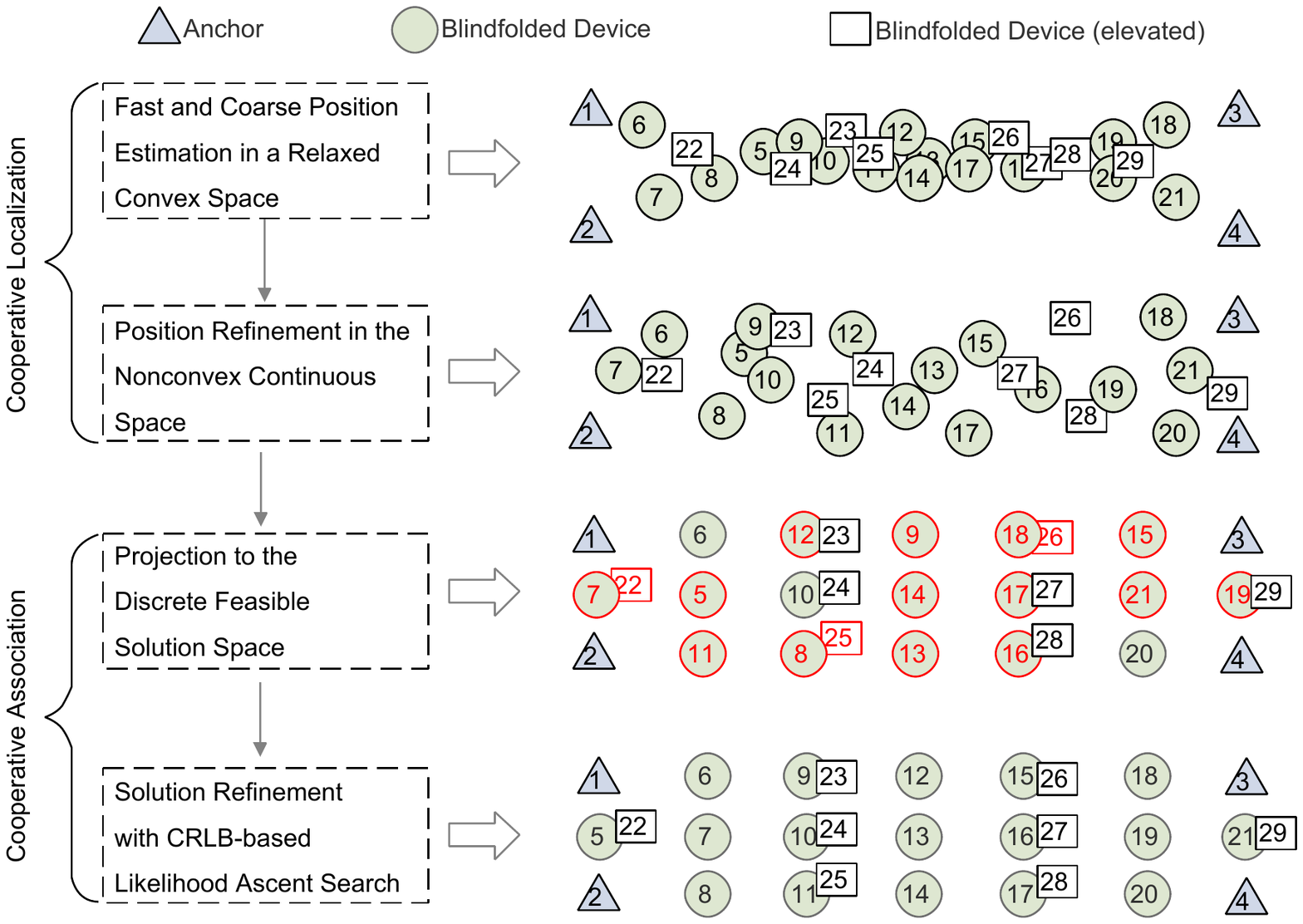}
			\captionof{figure}{The Overall Workflow of Our Approach}
			\label{fig:overall_process}
		\end{figure*}

		Fig.~\ref{fig:overall_process} provides the step of the proposed approach as well as the corresponding outcomes of the steps. A cooperative localization in continuous Euclidean space is conducted first by dropping Constraint (\ref{eqn:pos_constraint}). The resulting positions of the transmitters are then projected into the \textit{a priori} known discrete feasible solution space, followed by a likelihood ascent search in the discrete feasible solution space to refine the projection and association.
		
		\label{sec:algorithm:init}
		
		\subsection{Continuous 3D Space Cooperative Localization}
		\label{sec:continue_locn}
		By temporily dropping Constraint~(\ref{eqn:pos_constraint}), $\mathcal{P}_1$ reduces to a conventional RSS-based cooperative localization problem as
		\begin{align}
		\label{eqn:conventional_coop_locn}
		\mathcal{P}_2: &\\
		\nonumber
		\min_{\{\mathbf{p}_i\}_{i=1}^{N_t}}&\sum_{i=1}^{N_t} \sum_{j=i+1}^{N_t+N_a} \Big( P_0-10\gamma \log_{10} \frac{\|\mathbf{p}_i-\mathbf{p}_j\|}{d_0} - \tilde{R}_{ij} \Big)^2,
		\end{align}
		which can be solved efficiently with Gradient-based optimization methods.
		
		The performance of the gradient-based algorithm for $\mathcal{P}_2$ depends heavily on the initial point. We propose an initial positioning algorithm based on a second-order cone programming (SOCP) relaxation. First, we approximate $\mathcal{P}_2$ by the following min-max problem
		\begin{align}
		\label{eqn:min-max}
		\min_{\{\mathbf{p}_i\}_{i=1}^{N_t}} \max_{i=1}^{N_t}\max_{j=i+1}^{N_t+N_a} \left(P_0-10\gamma \log_{10} \frac{\|\mathbf{p}_i-\mathbf{p}_j\|}{d_0} - \tilde{R}_{ij} \right)^2,
		\end{align}
		which, by introducing a slacking variable $t$, can be reformulated as
		\begin{align}
		\mathcal{P}_3:
		\label{eqn:min-max-reformulated-objective}
		\min_{\{\mathbf{p}_i\}_{i=1}^{N_t}}& t\\
		\nonumber
		\text{s.t.} ~~~ & \left|P_0-10\gamma \log_{10} \frac{\|\mathbf{p}_i-\mathbf{p}_j\|}{d_0} - \tilde{R}_{ij}\right| \leq t \\
		\label{eqn:max_norm_constraint}
		&\forall i=1,...,N_t,~j=i+1,...,N_t+N_a,
		\end{align}
		By expanding and then simplifying (\ref{eqn:max_norm_constraint}), we obtain
		\begin{equation}
		d_0 \times 10^{\frac{P_0 - R^m_{ij} - t}{10\gamma}} \leq \|\mathbf{p}_i-\mathbf{p}_j\| \leq d_0 \times10^{\frac{P_0 - R^m_{ij} + t}{10\gamma}}
		\end{equation}
		
		For notiational simplicity, we let $t^{'} = 10^{\frac{t}{10\gamma}}$ and $\tilde{d}_{ij}= d_0 \times 10^{\frac{P_0 - R^m_{ij}}{10\gamma}}$. Then, we have
		
		\begin{equation}
		\label{eqn:inequality}
		\frac{\tilde{d}_{ij}}{t^{'}} \leq \|\mathbf{p}_i-\mathbf{p}_j\| \leq \tilde{d}_{ij} \times t^{'}
		\end{equation}
		
		As a result, (\ref{eqn:min-max-reformulated-objective}) can be transformed into the following optimization problem
		\begin{align}
		\label{eqn:init_est_inconvex}
		\mathcal{P}_4:~~~&
		\displaystyle \min_{t^{'}, \{\mathbf{p}_i\}_{i=1}^{N_t}} t^{'} \\
		& \textrm{s.t.} \quad \frac{\tilde{d}_{ij}}{t} \leq \|\mathbf{p}_i-\mathbf{p}_j\| \leq \tilde{d}_{ij}\times t^{'} \label{eqn:least_constraint} \\
		&~~~~~~~ i =1, \dots, N_t \quad j=i+1, \dots, N ~~\nonumber
		\end{align}
		where (\ref{eqn:least_constraint}) reflects the multiplicative ranging errors of RSS measurements. However, the left-hand side (LHS) of (\ref{eqn:inequality}) is a non-convex constraint. Therefore, we relax $\mathcal{P}_3$ by dropping the inequality constraint on the LHS of (\ref{eqn:inequality}), resulting in the following problem:
		\begin{eqnarray}
		\label{eqn:init_est_convex}
		& \displaystyle \min_{t^{'}, \{x_i, y_i, z_i\}_{i=1}^N} t^{'} \\
		& \textrm{s.t.} \quad \sqrt{(x_i - x_j)^2 + (y_i - y_j)^2 + (z_i - z_j)^2 } \leq \tilde{d}_{ij}t^{'} \nonumber \\
		& i =1, \dots, N_t \quad j=i+1, \dots, N \nonumber
		\end{eqnarray}
		The rationale of this convexification is two-fold: 1. The distance between two wireless devices can be arbitrarily small, but cannot exceed $\tilde{d}_{ij}t^{'}$; 2. The blindfolded devices' positions are arranged in the optimization such that $t^{'}$ is minimal.
		
		However, the solution to $\mathcal{P}_2$ violates (\ref{eqn:pos_constraint}) and therefore is not valid to $\mathcal{P}_1$. To obtain a valid permutation, we proceed to project the obtained solution to the discrete feasible solution space of $\mathcal{P}_1$ by associating the estimated location of each blindfolded device to one of the estimated locations of the transmitters, $\{\mathring{\mathbf{p}}_i\}_{i=1}^{N_t}$.
		
		\subsection{Discrete 3D Feasible Solution Projection}
		\label{sec:projection_to_feasible}
		
		By denoting the blindfolded devices' positions estimated in Section~\ref{sec:continue_locn} by $\widehat{\mathbf{p}}_i$, $i=1,...,N_t$, the projection into the discrete feasible solution space is achieved by:
		\begin{align}
		\label{eqn:projection}
		\min_{\mathbf{\pi}} & \sum_{i=1}^{N_t} \|\widehat{\mathbf{p}}_i-\mathring{\mathbf{p}}_{\mathbf{\pi}[i]}\| \\
		\textrm{s.t.} &~~~ \{\mathbf{\pi}[i]\}_{i=1}^{N_t} = \{1,...,N_t\},
		\end{align}
		where $\mathbf{\pi}$ is a vector describing the association between the continous positions estimated in the previous section and the known locations, with the $i$-th blindfolded device assigned to the $\mathbf{\pi}[i]$-th estimated continuous location,
		$\|\widehat{\mathbf{p}}_i-\mathring{\mathbf{p}}_{\mathbf{\pi}[i]}\|$ is the Euclidean distance between the estimated position of blindfolded device $i$ and the assigned known location.
		
		Now we have edge weights represented by Euclidean distances, the above problem is equivalent to the following linear assignment problem which can be solved with the Kuhn-Munkres algorithm:
		\begin{eqnarray}
		\label{eqn:kuhn_munkres}
		\mathcal{P}_5: ~~~ \min_{\mathbf{X}} & \sum_{i=1}^{N_t} \sum_{j=1}^{N_t} [\mathbf{C}]_{i,j} [\mathbf{X}]_{i,j} \\
		\textrm{s.t.} ~& [\mathbf{X}]_{i,j} \in \{0,1\}, \\
		&\sum_{i=1}^{N_t} [\mathbf{X}]_{i,j} = 1, \\
		&\sum_{j=1}^{N_t} [\mathbf{X}]_{i,j} = 1,
		\end{eqnarray}
		where $\mathbf{X}$ and $\mathbf{C}$ are $N_t$-dimensional square matrices, $[\mathbf{X}]_{i,j} = 1$ indicates that blindfolded device $i$ is assigned to the $j$-th known location, and
		$[\mathbf{C}]_{i,j}\triangleq \|\widehat{\mathbf{p}}_i-\mathring{\mathbf{p}}_j\|$.
		
		\subsection{Permutation and Refinement}
		The device-location assignment obtained in Section~\ref{sec:projection_to_feasible} is not optimal for the following reasons: (a) Constraint (\ref{eqn:pos_constraint}) is precluded in $\mathcal{P}_2$; (b) the solution to $\mathcal{P}_2$ is not guaranteed to be globally optimal; and (c) the projection in $\mathcal{P}_5$ does not preserve the optimality of $\mathcal{P}_2$.
		
		To tackle these issues, we adopt the Likelihood Ascent Search~(LAS)~\cite{5456045} algorithm to refine the device-location assignment through bounded local search in the feasible solution domain. The algorithm evaluates a candidate set of feasible solutions that can be derived from a given solution with specified transformations and choose the one with the lowest cost function (highest likelihood function) value. This process is conducted iteratively until no further improvement can be achieved.
		The detailed process of the LAS is provided in Alg. \ref{alg:sphere_decode}.
		\begin{algorithm}[H]
			\caption{Likelihood Ascent Search Algorithm}
			\label{alg:sphere_decode}
			\textbf{input}: Initial solution $\pi$ \\
			Set the best solution $\pi_b \leftarrow \pi $\\
			Set the lowest cost $E_b \leftarrow c(\pi_b)$ \label{line:residual_error} \\
			\Do{$E_b$ is updated}{
				Construct a candidate set $\mathcal{D}$ of feasible solutions that are close to $\pi_b$\\
				\For{each $\pi \in \mathcal{D}$}{
					\If {$c(\pi) < E_b$}{ \label{line:sphere_decode}
						$\pi_b \leftarrow \pi $\\
						$E_b \leftarrow c(\pi)$\\
					}
				}
			}
			\Return $\pi_b$
		\end{algorithm}
		A straightforward candidate set can be constructed by swapping the locations of $k~(k \geq 2)$ blindfolded devices, with $\mathcal{D}$ defined as
		\begin{align}
		\mathcal{D} = \left\{\pi \left| \sum_{i=1}^{N_t}\mathbbm{1}_{\pi_i \neq \pi_i^{(b)}} \leq k \right. \right\},
		\end{align}
		i.e. $\mathcal{D}$ contains all the permuations in the neighborhood of $\pi_b$ whose Hamming distances to $\pi_b$ are less than or equal to $k$.
		However, this approach has poor scalability since its time complexity is $O(|\mathcal{D}|)$, which increases exponentially with $k$.

		In the following section, we propose a CRLB-based likelihood ascent search algorithm, which significantly reduces the searching complexity while maintains high accuracy.
		
		\section{CRLB-based Likelihood Ascent Search}
		\label{sec:CRB_LAS}
		A key observation to the solution of $\mathcal{P}_5$ is that the errors in device-location assignment usually happen between devices that are geographically nearby. In other words, the devices are assigned to positions close to their true locations, but may be incorrectly swapped with their neighbors. This is one valuable property that we can leverage to perform dimensionality reduction by transforming $\mathcal{P}_1$ approximately to a series of low-dimensional problems that can be solved efficiently. Specifically, for each blindfolded device $i$, we solve
		\begin{align}
		\nonumber
		\mathcal{P}_{1,i}: ~~~ \min_{\pi} \quad c\big(\pi\big)\\
		\nonumber
		\text{s.t.} ~~ & \{\mathbf{p}_{\pi[k]}\}_{k \in \{i\}\bigcup \mathcal{N}_i}= \{\mathring{\mathbf{p}}_k\}_{k \in \{i\}\bigcup \mathcal{N}_i },\\
		& \mathbf{p}_{\pi[j]} = \mathring{\mathbf{p}}_j, \forall j \in \{i\}_{i=1}^{N_t}\texttt{\symbol{92}}\{k\}
		\end{align}
		where $\mathcal{N}_i$ is the set of neighbors of blindfolded device $i$. $\mathcal{N}_i$ contains indices which are close to the current blindfolded device and whose positions are most likely to be misplaced among their locations and the blindfolded device's location.\par
		
		We propose to accelerate the search process by only swapping the devices within $\mathcal{N}_i$. To achieve this, we derive the local 3D region of every installation position associated with the CRLB of the position. The CRLB provides the lower bound of the variance of an unbiased estimator and can provide a reliable and meaningful metric to specify the possible misplaced geographical neighbors in the Euclidean space of the devices.
		
		Define $\mathbf{p}=[x_1,...,x_{N_t}, y_1,...,y_{N_t}, z_1,...,z_{N_t}]^T$ which aggregates the actual x-, y-, and z-coordinates of all the blindfolded devices. Let $\widehat{\mathbf{p}}=[\widehat{x}_1,...,\widehat{x}_{N_t}, \widehat{y}_1,...,\widehat{y}_{N_t}, \widehat{z}_1,...,\widehat{z}_{N_t}]^T$ denote the corresponding estimated coordinates obtained by solving $\mathcal{P}_2$. According to the CRLB, we have
		\begin{align}
		\mathbb{E}[(\widehat{\mathbf{p}}-\mathbf{p})(\widehat{\mathbf{p}}-\mathbf{p})^T]\ge
		\mathbf{F}^{-1},
		\end{align}
		where $\mathbf{F}$ is the Fisher information matrix (FIM), as given by
		\begin{eqnarray}
		\label{eqn:fim}
		\mathbf{F} = \left[ \begin{array}{lll}
		\mathbf{F}_{xx} & \mathbf{F}_{xy} & \mathbf{F}_{xz} \\
		\mathbf{F}_{xy}^T & \mathbf{F}_{yy} & \mathbf{F}_{yz} \\
		\mathbf{F}_{zx}^T & \mathbf{F}_{yz}^T & \mathbf{F}_{zz}
		\end{array} \right]
		\end{eqnarray}
		Each sub-matrix of $\mathbf{F}$ is defined by
		\begin{eqnarray}
		\label{eqn:fisher_information}
		\left[F_{xx}\right]_{k,l} &=& \left\{ \begin{array}{ll}
		\gamma_c \sum_{i=1}^{N_t} \frac{(x_k - x_i)^2}{d_{k,i}^4} ~~~~~~~~ &,~\text{if}~ k=l \nonumber \\
		-\gamma_c \frac{(x_k - x_i)^2}{d_{k,i}^4} &,~\text{if}~k \neq l \nonumber
		\end{array} \right. \\
		\left[F_{xy}\right]_{k,l} &=& \left\{ \begin{array}{ll}
		\gamma_c \sum_{i=1}^{N_t} \frac{(x_k - x_i)(y_k - y_i))}{d_{k,i}^4}& ,~\text{if}~k=l \nonumber \\
		-\gamma_c \frac{(x_k - x_i)(y_k - y_i)}{d_{k,i}^4}& ,~\text{if}~k \neq l \nonumber
		\end{array} \right. \\
		\end{eqnarray}
		where $\gamma_c = (\frac{10\gamma}{\sigma \log{10}})^2$, 
		and $d_{k,i}$ is the actual Euclidean distance between devices $k$ and $i$~($i\neq k$). The remaining sub-matrices of $\mathbf{F}$ can be easily obtained by substituting $x$ and $y$ with corresponding coordinate symbols.

		Therefore, for each blindfolded device $i$, we have

		\begin{align}
		\mathbb{E}[&(\widehat{\mathbf{p}}_i-\mathbf{p}_i)(\widehat{\mathbf{p}}_i-\mathbf{p}_i)^T]\succcurlyeq \\
		\nonumber
		& \begin{bmatrix*}[l]
		(\textbf{F}^{-1})_{i,i} & (\textbf{F}^{-1})_{i,i+N_t} & (\textbf{F}^{-1})_{i,i+2N_t} \\
		(\textbf{F}^{-1})_{i+N_t,i} & (\textbf{F}^{-1})_{i+N_t,i+N_t} & (\textbf{F}^{-1})_{i,i+2N_t} \\
		(\textbf{F}^{-1})_{i+2N_t,i} & (\textbf{F}^{-1})_{i+2N_t,i} & (\textbf{F}^{-1})_{i+2N_t,i+2N_t}
		\end{bmatrix*}
		\end{align}
		We define the neighbors of any blindfolded device $i$ as those devices which lie within the 95\% confidence region of $\widehat{\mathbf{p}}_i$, i.e.,
		\begin{align}
		\mathcal{N}_i \triangleq \{j|
		(
		\mathring{\mathbf{p}}_{\mathbf{\pi}[i]}
		-
		\mathring{\mathbf{p}}_{\mathbf{\pi}[j]}
		)^{T}
		\Sigma^{-1}
		(
		\mathring{\mathbf{p}}_{\mathbf{\pi}[i]}
		-
		\mathring{\mathbf{p}}_{\mathbf{\pi}[j]}
		) \leq \chi^{2}_3(0.95)\}
		\end{align}
		where $ \chi _{3}^{2}(\cdot)$ is the quintile function of the Chi-squared distribution with three degrees of freedom.
		
		We now run the LAS algorithm in such a way that only the devices within the CRLB region of a blindfoled device with $p=95\%$ confidence are tested. Therefore, $D_n$ can be specified by
		\begin{align}
		D_n = \{\pi|\pi_k = S_b[k],~ \forall k \notin \{i\} \bigcup \mathcal{N}_i\}.
		\end{align}
		
		Since the continuous space search usually produces solutions with errors mostly between neighbors, such a space specified by $\mathcal{N}_i$ allows for a low-dimensional search to balance the time complexity of the algorithm and the accuracy of the result.
		
		\section{Experiment Results}
		\label{sec:simulation}
		The performance of the proposed approach is evaluated through extensive computer simulations and lab tests, and compared against well-known metaheuristics that have been extensively used to solve intractable combinatorial optimization problems. The following approaches are considered:
		
		\begin{itemize}
			\item 	\textbf{CRLB-LAS}:  The approach described in Fig.~\ref{fig:overall_process}, using the CRLB-based LAS algorithm described in Section~\ref{sec:CRB_LAS}.
			
			\item  \textbf{Continuous Space Optimization(CSO)}: The algorithm of  continuous space cooperative localization followed by projection into the feasible solution space, as described in Section~\ref{sec:algorithm:init}.
			
			\item 	\textbf{Simulated Annealing~(SA)}~\cite{Kirkpatrick671}: A probabilistic approach often used for optimization in a large discrete space(e.g., assignment problem)~\cite{7580628,1275542}. SA mimics annealing processes in metallurgy, which is a technique involving heating and controlled cooling of a material. In the experiments, we utilize parameterized sigmoid as the annealing function for modeling the relationship between the annealing temperature and the searching scope. Exploration is conducted by randomly reverting large slices of the current best solution in the first stage, and then by searching in the neighborhood of the current best solution through randomly reverting small slices of the solution. Multiple sigmoid functions are tried, and the one with the best performance is used.

			\item \textbf{Genetic Algorithm~(GA)}~\cite{4358780,7289449}: A general-purpose problem-solving approach that is based on the Darwinian principle of reproduction and survival of the fittest and analogs of naturally occurring genetic operations such as selection, crossover, and mutation, which has shown its strength on many intractable optimization problems including TSP~\cite{Razali2011GeneticAP, 10.1007/3-540-45724-0_28,4067083} and assignment problem~\cite{7173042,1042252,1306435,5530324}. GA is conducted in the experiments using 100,000 generations, 50 populations, and 10 elites.
			
			\item \textbf{Tabu search~(Tabu)}~ \cite{GLOVER1986533,6305447,7283646}: A metaheuristic for discrete optimization by employing local search methods, which improves the performance by accepting worsening solutions if no improvement is available. Besides, prohibitions (hence the term tabu) are introduced to divert the search from past solutions. In the experiments, we set the memory size of Tabu to 50, i.e., during the search Tabu will not step into the last 50 swaps that it has tried in order to improve its exploration capability.
			
		\end{itemize}

		The experiments were carried out on a laptop with Intel Core i5-6300 CPU@2.4GHz and 16GB memory. The simulation results are averaged over one hundred independent executions.
		
		\subsection{Smart Airplane Cabin}
		\label{sec:simu_airplane}
\begin{figure*}[tb]
			\centering
			\includegraphics[width=6in]{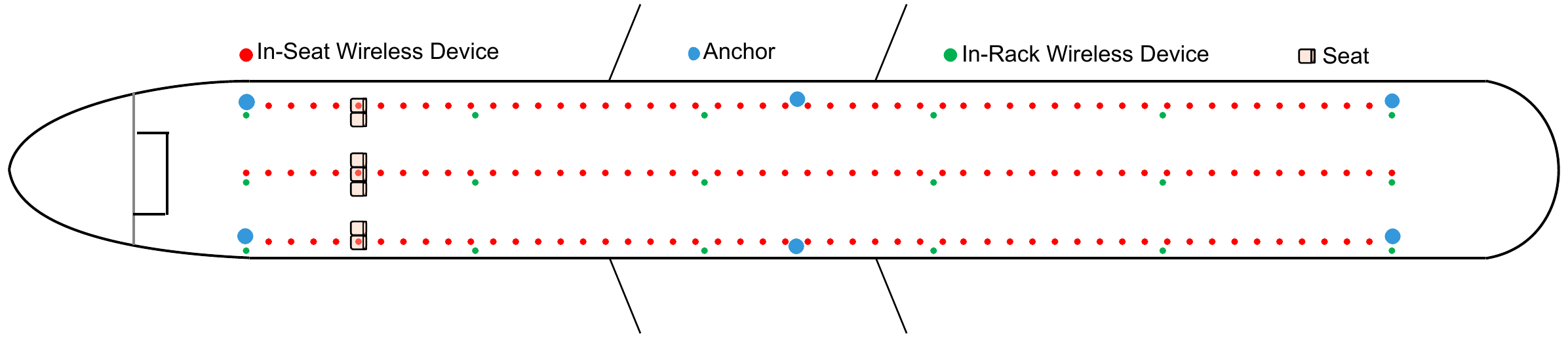}
			\caption{Layout of the airplane cabin and locations of wireless devices in the simulation (top view). }
			\label{fig:position_in_cabin}
		\end{figure*}

		We study the performance of the considered approaches in a smart airplane cabin by computer simulations.				
		We use a 35.7m$\times$4.72m$\times$1.5m cuboid to emulate a typical airplane cabin. The cabin has three sub-rows in each row of seats, and the gap between
		sub-rows is 2.36 m. Among each sub-row, the seats are 0.7 m apart. On
		the back of one seat in each sub-row, a wireless device is installed. There are also multiple wireless devices fitted in the overhead luggage rack. Six devices are used as anchors with their locations shown in Fig.~\ref{fig:position_in_cabin}.
		
		Fig. \ref{fig:swlas} shows the accuracy of CRLB-LAS and CSO under different standard deviation values of the shadow fading (i.e., $\sigma$), where the results are obtained over 100 independent experiments. It is shown that CRLB-LAS significantly outperforms CSO. Particularly, the medium accuracy of CRLB-LAS is 100\% for $\sigma<4$~dB, while that of CSO is between 92\% and 96\%. The accuracy degrades as $\sigma$ increases for both approaches, but a medium accuracy of 89-98\% is achieved by CRLB-LAS when $\sigma$ is between $5$~dB and $7$~dB. In contrast, the medium accuracy of CSO lies between 81\% and 89\% when $\sigma$ increases from 5dB to 7dB. Overall, CRLB-LAS outperforms CSO by 7 -- 9\%.\par
		
		\begin{Figure}
			\centering
			\includegraphics[width=3.3in, height=2.8in]{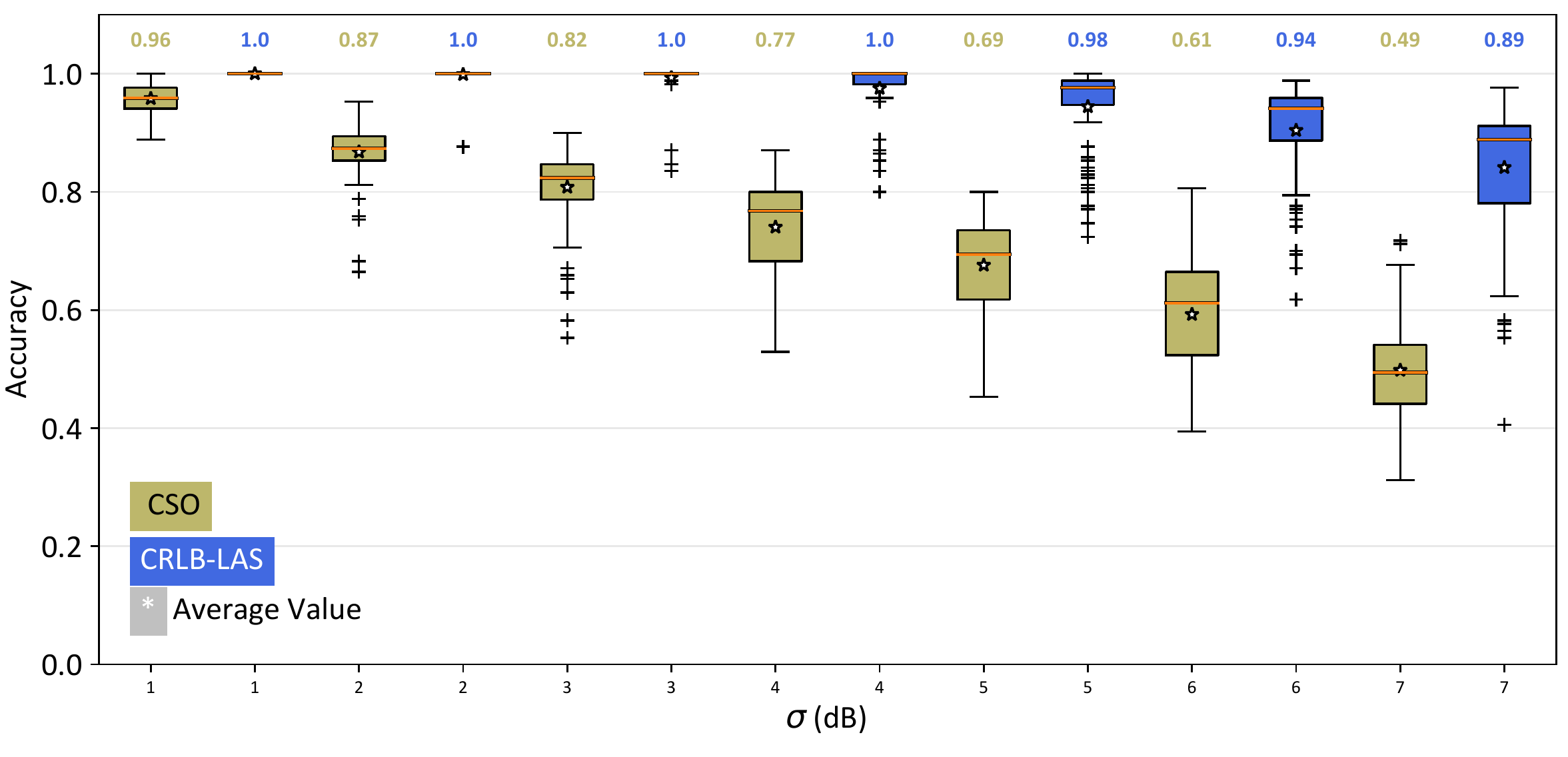}
			\captionof{figure}{The accuracy of CRLB-LAS and CSO under different noise strengths}
			\label{fig:swlas}
		\end{Figure}
		
		Fig.~\ref{fig:approach_compare} compares the median accuracy between CRLB-LAS and the state-of-the-art metaheuristics. It can be seen that the accuracies of Anneal and GA are close to zero, due to the extremely high dimension of the solution space. Tabu performs substantially better than Anneal and GA thanks to its efficient exploration strategy. CRLB-LAS significantly outperforms all of the metaheuristics, achieving close-to-100\% accuracy.
		
		\begin{Figure}
			\centering
			\includegraphics[width=3.7in]{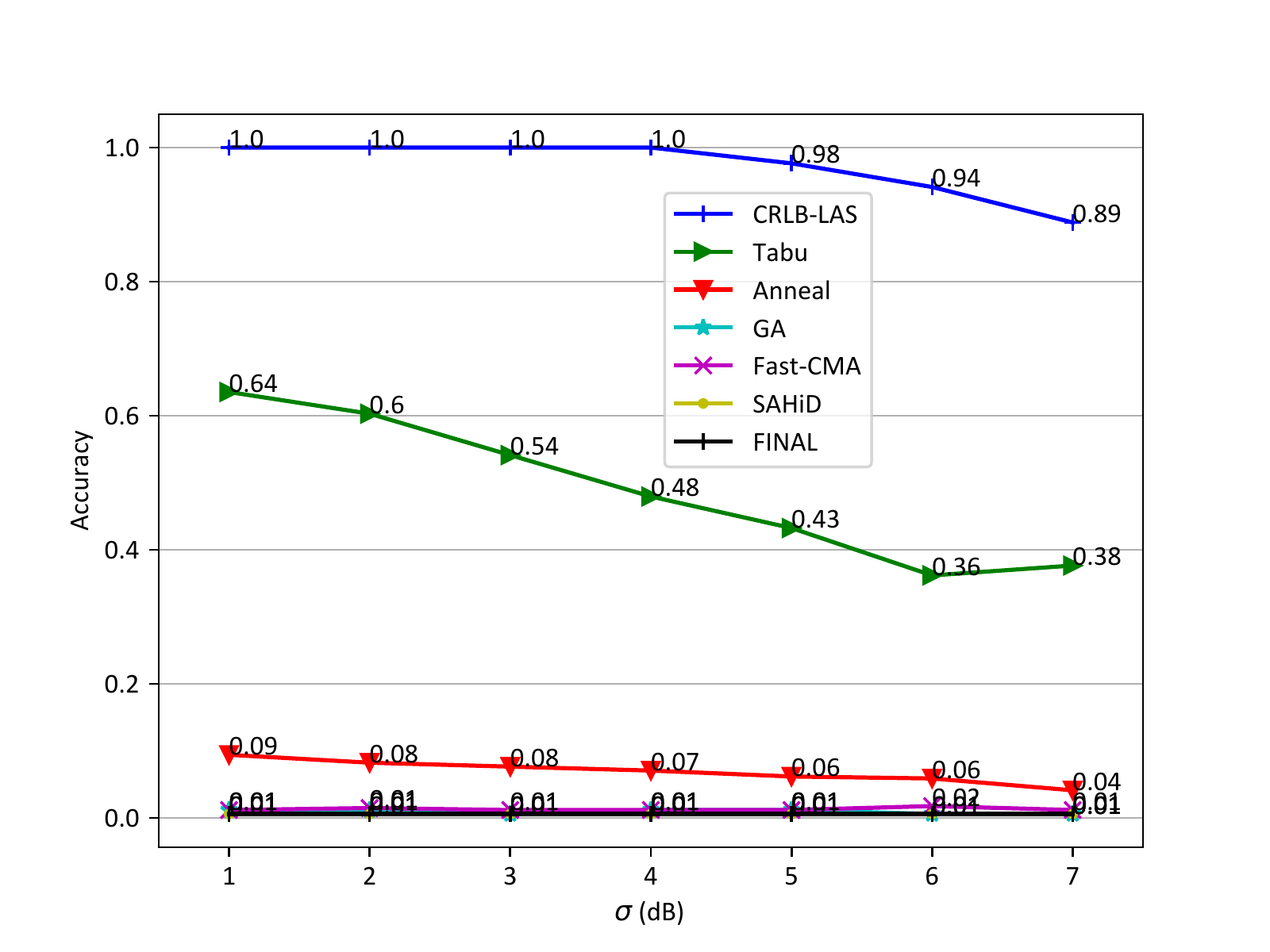}
			\captionof{figure}{Median accuracy of the considered approaches in airplane cabin.}
			\label{fig:approach_compare}
		\end{Figure}
		
		\begin{Figure}
			\centering
			\includegraphics[width=3.4in]{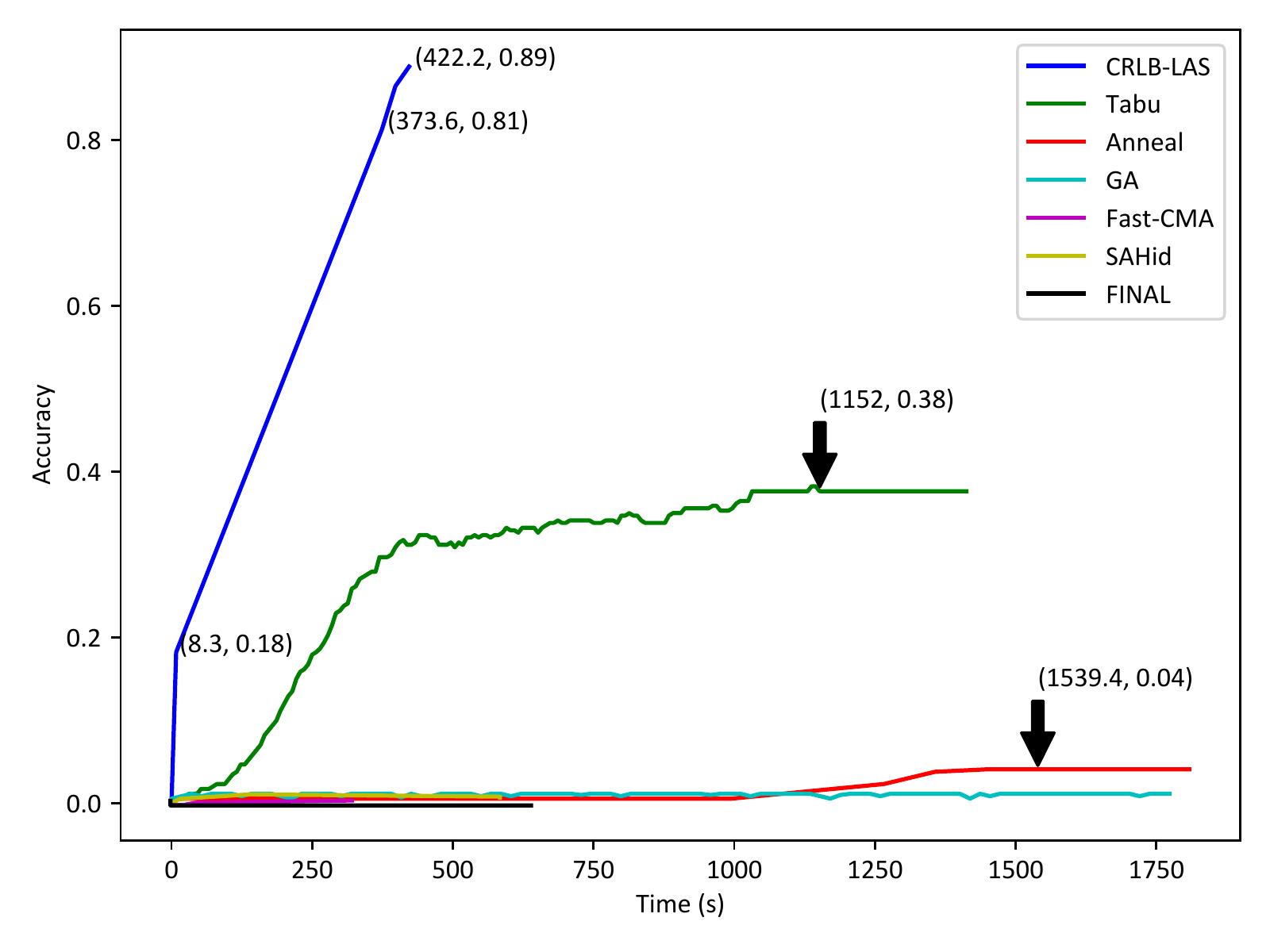}
			\captionof{figure}{The change of accuracy with execution time for considered algorithms ($\sigma = 7$dB).}
			\label{fig:compare_performance}
		\end{Figure}
		Fig.~\ref{fig:compare_performance} shows the changing accuracy with execution time for the considered algorithms. Each metaheuristic is carefully tuned for the best performance of the considered problem. It can be seen that CRLB-LAS converges much faster than the others, and converges within 422.2 seconds on average, while Tabu and Anneal converge in 1,152 and 1,539.4 seconds, respectively, which shows the importance of the continuous space search of CRLB-LAS. From the output of the continuous space search, it is much easier to reach the optimal solution than exploring in the vast solution space. While GA converges faster than CRLB-LAS, it clearly has much poorer accuracy. Although GA performs well on many NP-hard problems such as TSP, in this case GA is trapped at a local optimum even though the maximum generation is set to 100,000.

		\subsection{Experiments with COTS WiFi Devices}
		We have conducted experiments with COTS WiFi devices to evaluate the performance of the proposed technique.
		A testbed based on the Cypress CYW54907 WiFi system-on-chip is developed. Each device is fitted with a vertically orientated, circularly polarised antenna, which provides a near omni-directional sensitivity with only two narrow beamwidth minima directly above and below the antenna. Fig.~\ref{fig:Cypress-CYW54907} shows the device.
		
		\begin{Figure}
			\centering
			\includegraphics[width=3in]{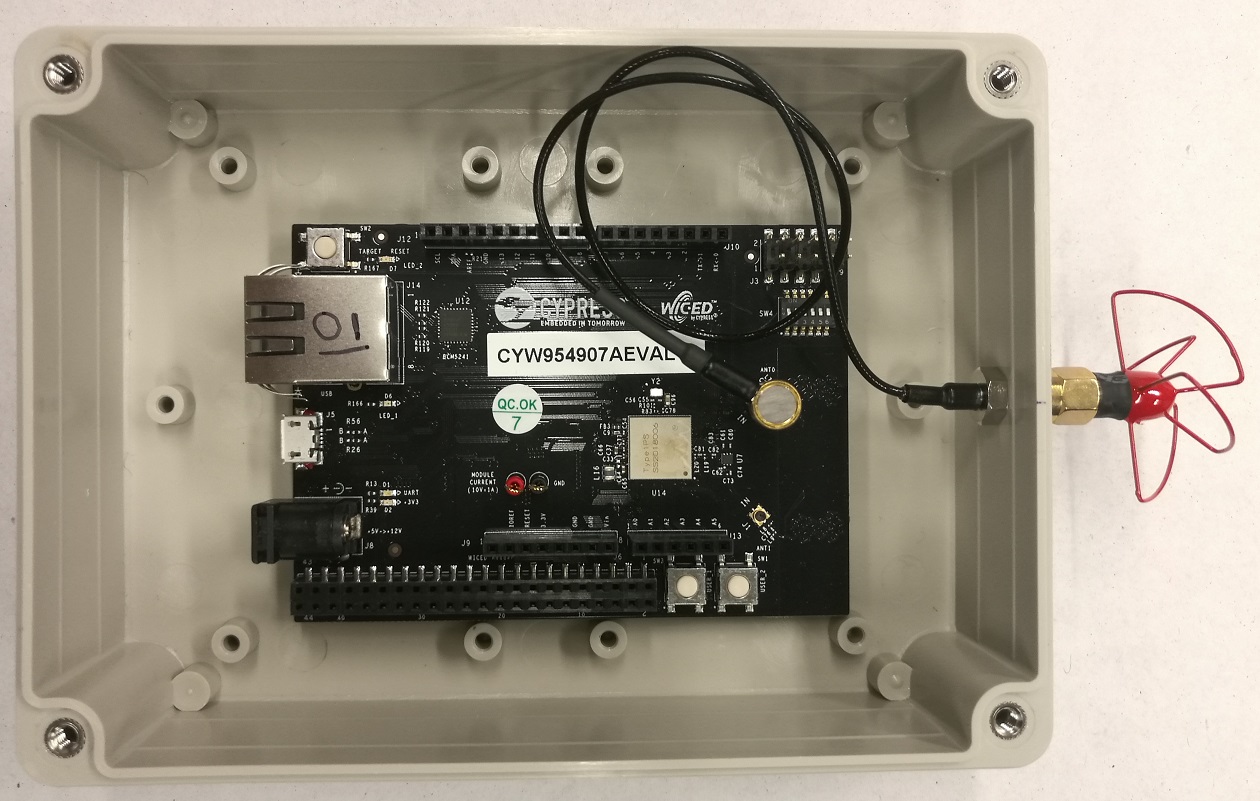}
			\captionof{figure}{Cypress CYW54907}
			\label{fig:Cypress-CYW54907}
		\end{Figure}
		\begin{Figure}
			\centering
			\includegraphics[width=3in]{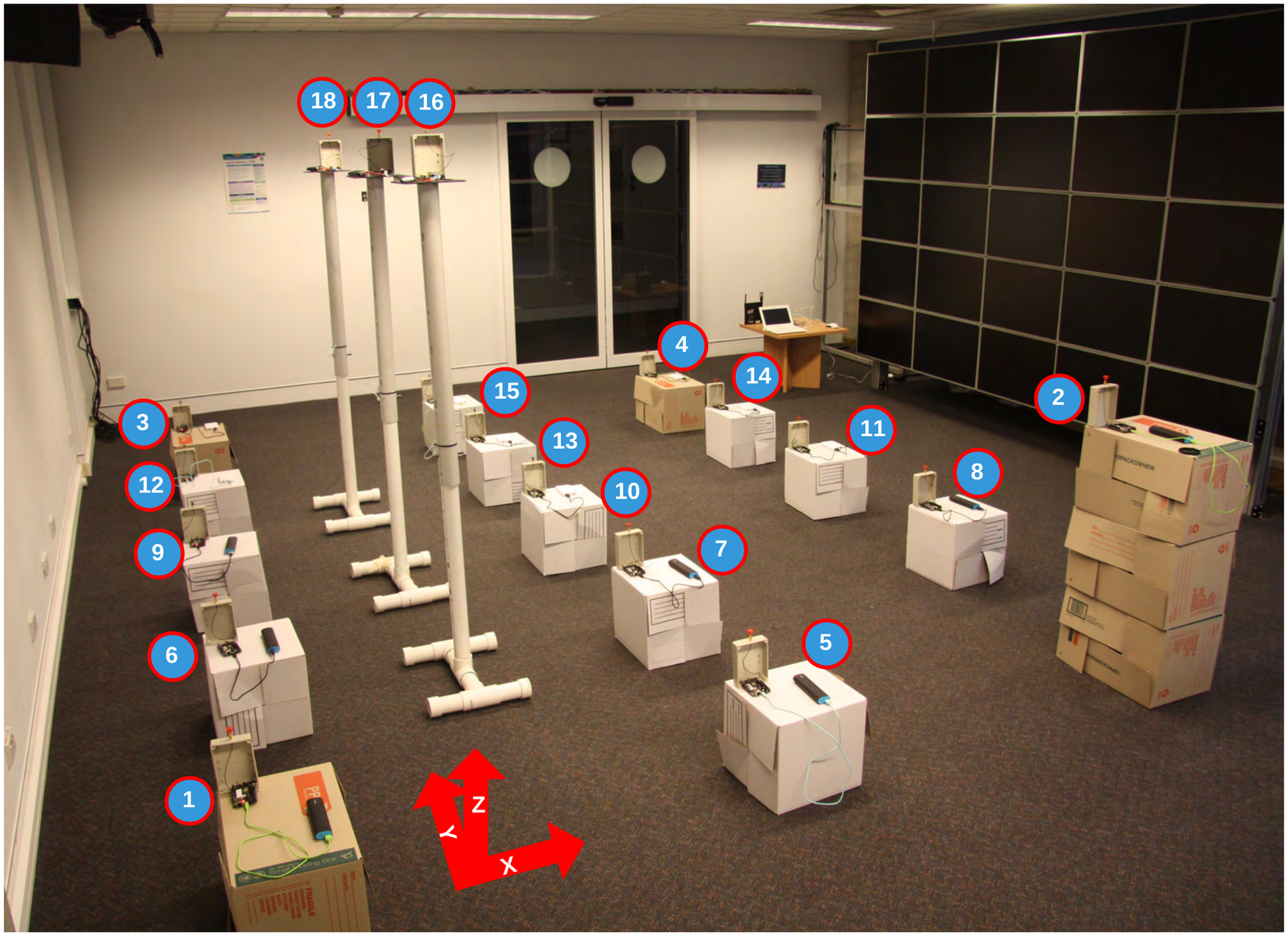}
			\captionof{figure}{Experiment Setup.}
			\label{fig:experiment}
		\end{Figure}
		
		A network of the WiFi system-on-chip devices is deployed in each experiment. Each device is configured as both a WiFi access point and a WiFi station.
		The access point in each device broadcasts beacon frames every 100 milliseconds, while the station scans for beacon frames from nearby devices and records the received signal strength. The collected data is transmitted to a laptop computer periodically.
		
		The devices are configured to hop across multiple WiFi channels.
		The measured signal strengths are averaged over all channels. The layout of the WiFi devices during the experiments is shown in Fig.~\ref{fig:experiment}. The devices on top of the four brown boxes at the corners are used as anchors with known locations and heights 48cm, 48cm, 48cm, and 110cm. There are 10 devices on top of white boxes with height 48cm and three devices on stands with height 147cm/195cm giving 13 blindfolded devices. The possible coordinates of all WiFi devices are listed in Table~\ref{tab:coordinate_of_devices}.

		\begin{center}
			\captionof{table}{Coordinates of WiFi devices during the experiments. Device Types: A = Anchor, B = Blindfolded device.}
			\label{tab:coordinate_of_devices}
			\begin{tabular}{c|c|c|c|c}
				No. & \textbf{X (m)} & \textbf{Y (m)} & \textbf{Z (m)} &\textbf{Device Type}\\
				\hline
				1& 0 & 1 &0.48 &A  \\
				2& 3 & 1 &1.1 & A  \\
				3& 0 & 5 &0.48 & A\\
				4& 3 & 5 &0.48 & A \\
				5& 1.5 & 1 &0.48 & B \\
				6& 0 & 2 &0.48 & B \\
				7& 1.5 & 2 &0.48 &  B \\
				8& 3 & 2 &0.48 & B \\
				9& 0 & 3 &0.48 & B \\
				10& 1.5 & 3 &0.48 & B \\
				11& 3 & 3 &0.48 &  B \\
				12 &0 & 4 &0.48 &  B \\
				13 & 1.5 & 4 &0.48 &  B \\
				14 & 3 & 4 &0.48 & B \\
				15 & 1.5 & 5 &0.48 & B \\
				16 & 0.75 & 2 &1.47 or 1.97 &  B \\
				17& 0.75 & 3 &1.47 or 1.97  & B \\
				18& 0.75 & 4&1.47 or 1.97  &  B \\
				\hline
			\end{tabular}
		\end{center}
		
		Fig.~\ref{fig:PathLossModel} shows the relationship between the measured signal strength and the corresponding distance and a path loss model fitted to the measurement data, which are generated from the measurement results and the ground truth of the device locations. The values of $\gamma$ and $p_0$ in (\ref{eqn:rss_ideal}) are 1.53 and -42.35, respectively.
		\begin{Figure}
			\centering
			\includegraphics[width=3.5in]{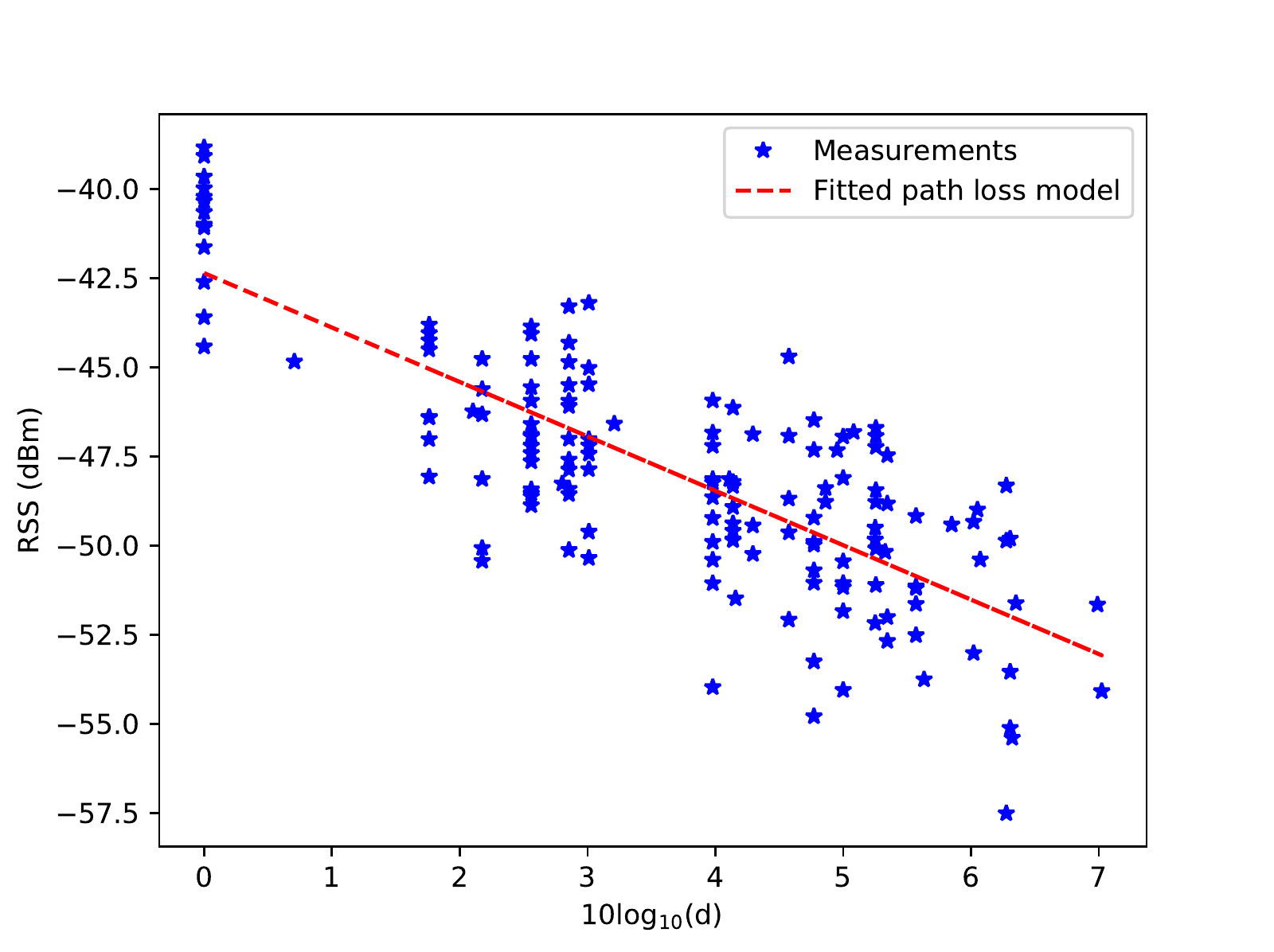}
			\captionof{figure}{The relationship between signal strength and distance.}
			\label{fig:PathLossModel}
		\end{Figure}

		Fig.~\ref{fig:lab_acc_137} shows the accuracies of the considered approaches over 100 runs. Each of the runs is independently randomly initialized. It is seen that CRLB-LAS achieved 100\% accuracy, which is significantly better than the rest of the techniques. The performance of metaheuristics depends heavily on the random initialization. The likelihood of achieving 100\% accuracy is about 41\% and 59\% for Tabu and GA, respectively, when the heights of devices 16, 17 and 18 are 1.47m.
The likelihood is about 56\% and 70\% for Tabu and GA, respectively, when the heights of the devices are 1.97m.
The highest accuracy achieved by Anneal is about 65\%.

		\begin{Figure}
			\centering
			\includegraphics[width=3.5in]{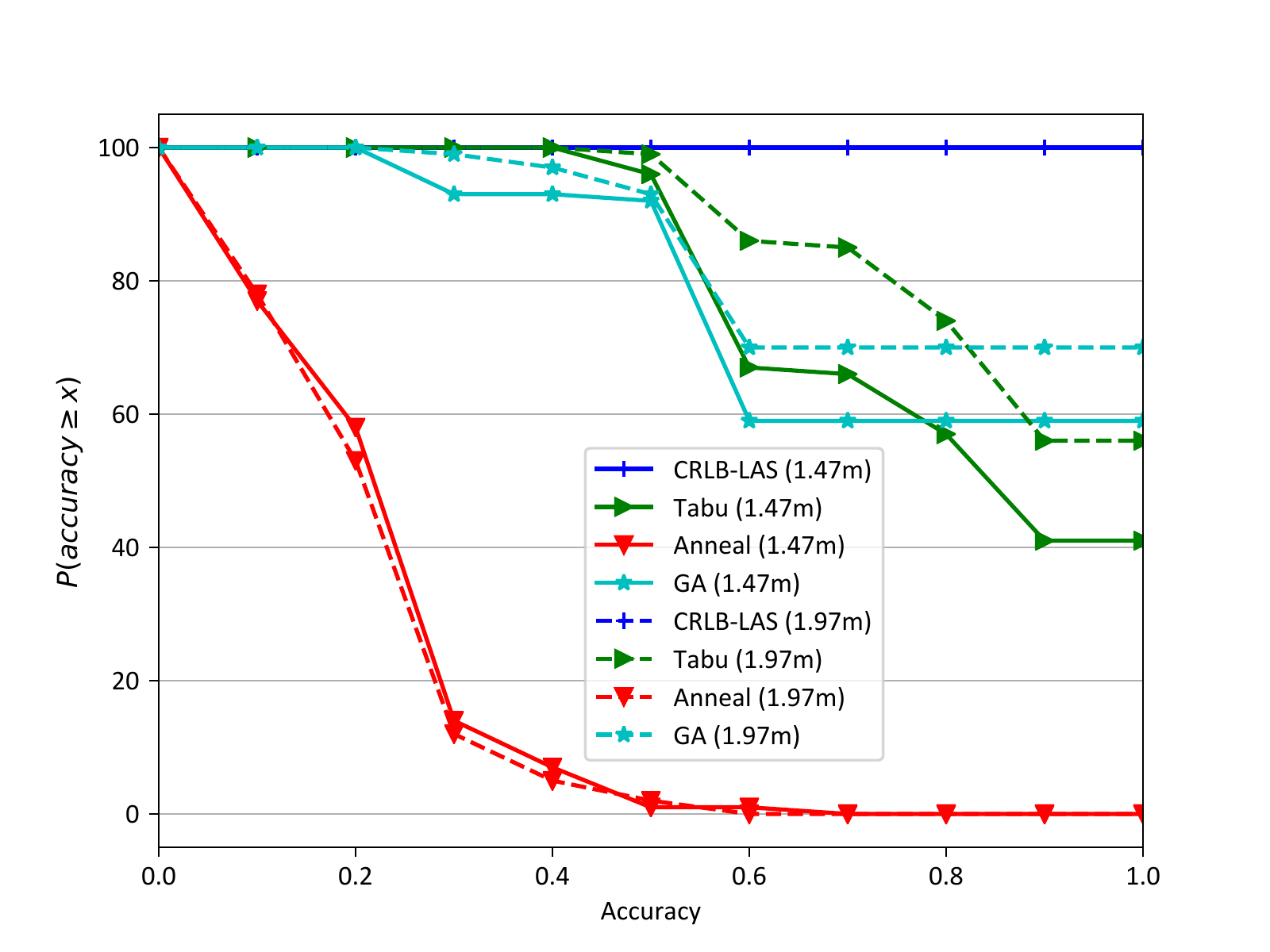}
			\captionof{figure}{The accuracies of the considered approaches with independent random initializations.}
			\label{fig:lab_acc_137}
		\end{Figure}
		
		Fig.~\ref{fig:lab_runtime} depicts the changing accuracy of the considered approaches with the growth of the execution time in the experiments. It can be seen that CRLB-LAS achieves 100\% accuracy in less than 1.5 seconds. In contrast, it takes more than 8 seconds for Anneal and GA to converge, with the corresponding average accuracy of around only 20\% and 50\%, respectively. Tabu converges faster than the other approaches; and its final average accuracy is about 85\%, which is lower than CRLB-LAS. For large-scale systems, the performance of Tabu is significantly lower than CRLB-LAS, as can be seen in Fig.~\ref{fig:compare_performance}.
		
		\begin{Figure}
			\centering
			\includegraphics[width=3.5in]{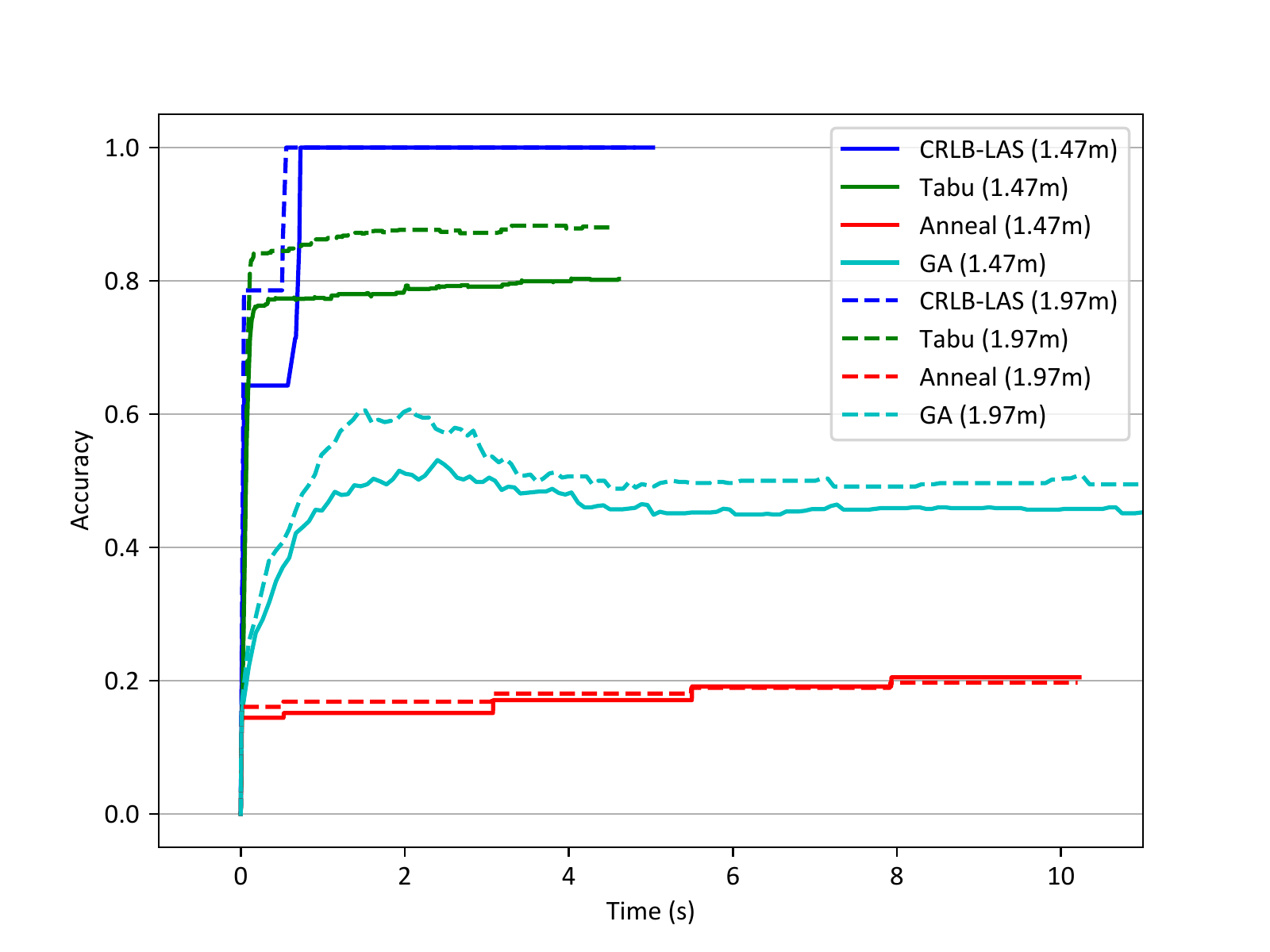}
			\captionof{figure}{The accuracies of the different approaches with the growing execution time in the experiments. The average accuracy of GA converges to between 0.45 and 0.5 around 745 seconds which is far off to the right of this figure and hence suppressed for the clarity of the figure.}
			\label{fig:lab_runtime}
		\end{Figure}

		\subsection{Smart Building}
		We also investigate the performance of CRLB-LAS and the metaheuristics in a smart building scenario, where multiple detectors are installed to enhance the safety of the building. The floorplan of the building is shown in Fig.~\ref{fig:building_simulate}. There are 42 detectors installed in the 19.5m$\times$19m$\times$2.6m space, including cameras, glass-break detectors, motion detectors, gas detectors, etc. Some detectors (e.g., Cameras) are installed near the ceiling of the building, while the others are installed close to the floor (e.g., gas detectors). There are four anchors installed at the corners.
		
		\begin{Figure}
			\centering
			\includegraphics[width=3in]{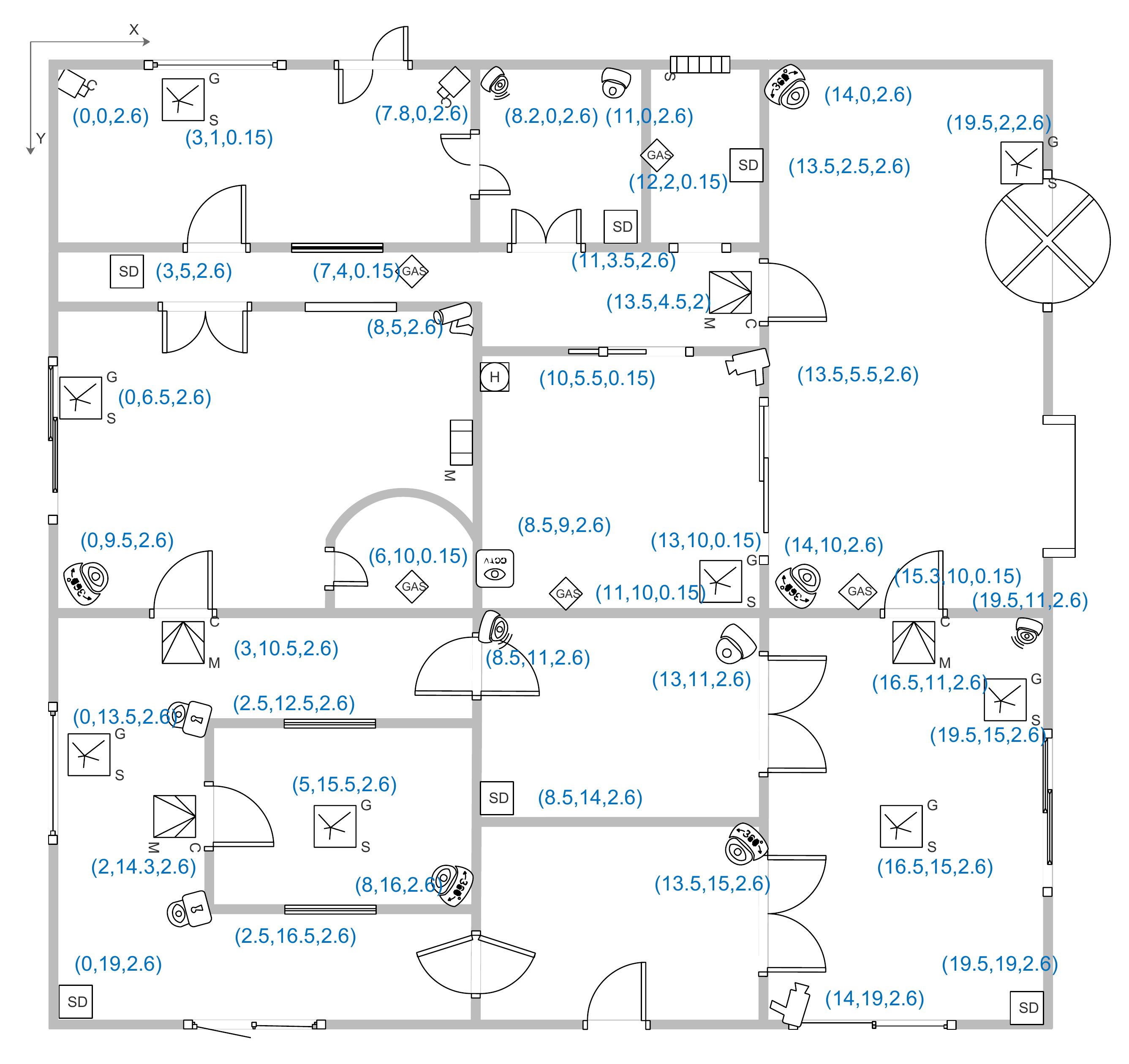}
			\captionof{figure}{Simulation setup of building security access.}
			\label{fig:building_simulate}
		\end{Figure}

		Fig.~\ref{fig:building_simulation} shows the median accuracies of the considered approaches. We can see that CRLB-LAS still outperforms the others significantly.
		\begin{Figure}
			\centering
			\includegraphics[width=3in]{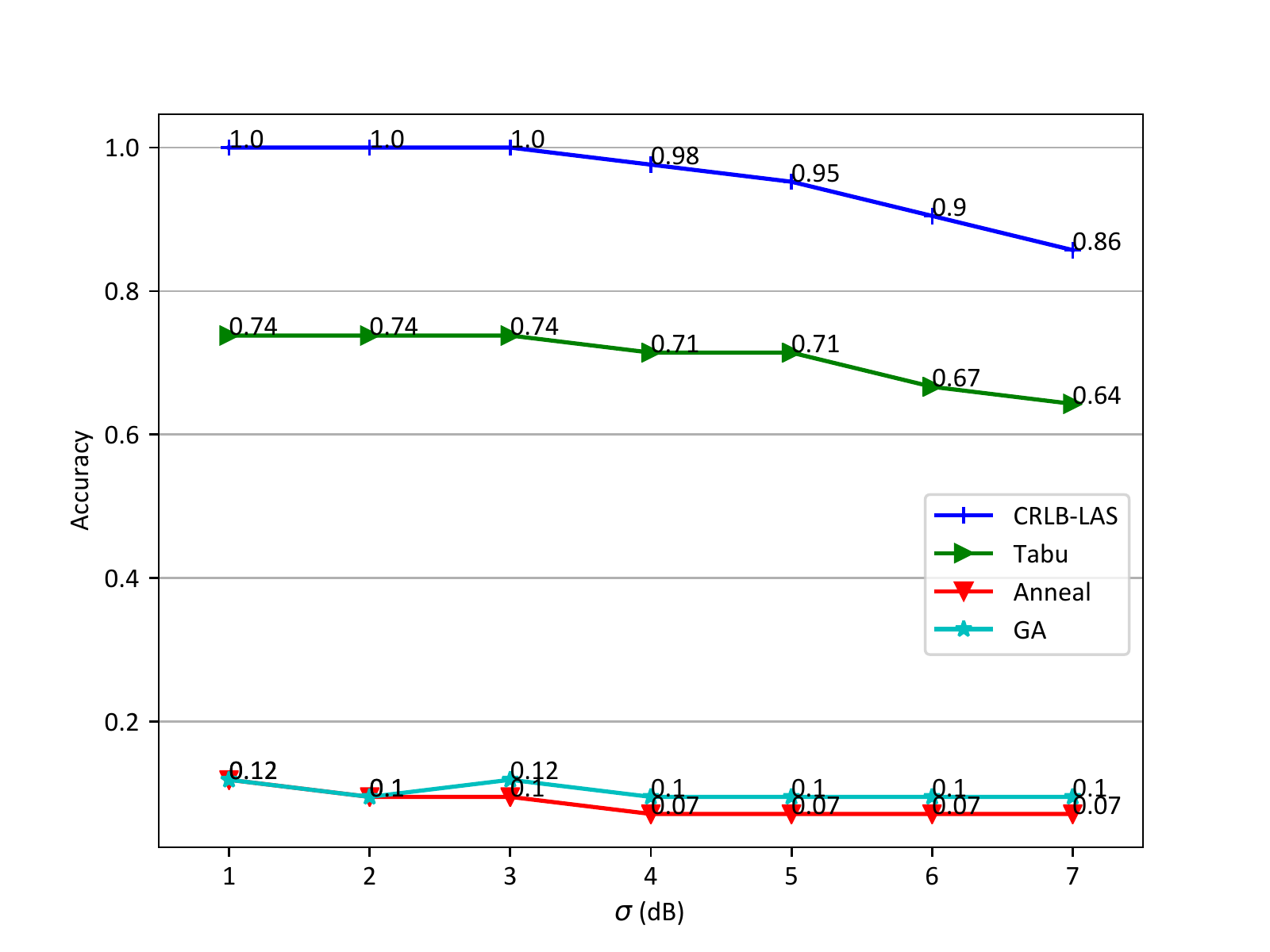}
			\captionof{figure}{Accuracies in smart building scenario.}
			\label{fig:building_simulation}
		\end{Figure}
		
		\section{Conclusion}
		\label{sec:conclusion}
		The association of a large number of low-cost Internet-of-Things (IoT) sensors and their possible installation locations is studied in this paper. An efficient approach to solve the corresponding permutation combinatorial optimization problem is proposed. The approach integrates continuous space cooperative localization and permutation space likelihood ascent search. We evaluate the performance of the proposed approach with extensive experiments, showing that the proposed approach significantly outperforms state-of-the-art combinatorial optimization algorithms and achieves close-to 100\% accuracy with affordable execution time.
		\bibliographystyle{IEEEtran}

		\bibliography{myref}
		
		%
		%
		%
	\end{multicols}
\end{document}